\documentclass[acmsmall]{acmart}
\AtBeginDocument{%
  }

\usepackage{array}
\usepackage{tabularx}
\usepackage{multirow}
\usepackage{float}
\usepackage{footnote}
\usepackage{url}
\usepackage{hyperref}
\usepackage{rotating}
\usepackage{graphicx}
\usepackage{booktabs} 
\usepackage{balance}
\usepackage{tabulary}
\usepackage{tabularray}
\usepackage{longtable}
\usepackage{color}
\usepackage{amsmath}
\usepackage{xcolor}
\usepackage{mathtools,tikz,caption}
\usepackage[framemethod=TikZ]{mdframed}
\usepackage{xcolor}
\usepackage[skins,breakable,most]{tcolorbox}
\usepackage{algorithm}
\usepackage{algpseudocode}
\usepackage{caption}
\usepackage{subcaption}
\usepackage{adjustbox}
\usepackage{wrapfig}
\usepackage{amsmath} 
\usepackage{wrapfig}
\newcommand{\miryung}[1]{{{\color{blue}{MK: #1}}}}
\newcommand{\mycomment}[1]{}

\newcommand{\tool}{{\it Soft Assertion Fuzzer }}
\providecommand{\keywords}[1]{\textbf{\textit{Keywords--}} #1}

\setcopyright{cc}
\setcctype{by}
\acmDOI{10.1145/3729394}
\acmYear{2025}
\acmJournal{PACMSE}
\acmVolume{2}
\acmNumber{FSE}
\acmArticle{FSE124}
\acmMonth{7}
\received{2025-02-25}
\received[accepted]{2025-04-01}

\begin{document}

\title{Automatically Detecting Numerical Instability in Machine Learning Applications via Soft Assertions}


\author{Shaila Sharmin}
\authornote{Both authors contributed equally to this research.}
\email{ssharmin@iastate.edu}
\orcid{0009-0009-0854-5404}
\affiliation{%
  \institution{Iowa State University}
  \city{Ames}
  % \state{Iowa}
  \country{USA}
}

\author{Anwar Hossain Zahid}
\authornotemark[1]
\email{ahzahid@iastate.edu}
\orcid{0000-0002-7014-112X}
\affiliation{%
  \institution{Iowa State University}
  \city{Ames}
  % \state{Iowa}
  \country{USA}
}

\author{Subhankar Bhattacharjee}
\orcid{0009-0001-5976-705X}
\affiliation{%
  \institution{Iowa State University}
  \city{Ames}
  \country{USA}
}
\email{s7bhat@iastate.edu}

\author{Chiamaka Igwilo}
\orcid{0009-0005-7096-6753}
\affiliation{%
  \institution{Iowa State University}
  \city{Ames}
  \country{USA}
}
\email{Igwilo@iastate.edu}

\author{Miryung Kim}
\orcid{0000-0003-3802-1512}
\affiliation{%
  \institution{University of California at Los Angeles}
  \city{Los Angeles}
  \country{USA}
}
\email{miryung@cs.ucla.edu}

\author{Wei Le}
\orcid{0000-0002-6797-0648}
\affiliation{%
  \institution{Iowa State University}
  \city{Ames}
  \country{USA}
}
\email{weile@iastate.edu}




% %%
% %% Keywords. The author(s) should pick words that accurately describe
% %% the work being presented. Separate the keywords with commas.
% \keywords{Do, Not, Us, This, Code, Put, the, Correct, Terms, for,
% %% A "teaser" image appears between the author and affiliation
% %% information and the body of the document, and typically spans the



\begin{abstract}
Machine learning (ML) applications have become an integral part of our lives. ML applications extensively use floating-point computation and involve very large/small numbers; thus, maintaining the numerical stability of such complex computations remains an important challenge. Numerical bugs can lead to system crashes, incorrect output, and wasted computing resources. In this paper, we introduce a novel idea, namely \textit{soft assertions (SA)}, to encode safety/error conditions for the places where numerical instability can occur. A soft assertion is an ML model automatically trained using the dataset obtained during unit testing of unstable functions. Given the values at the unstable function in an ML application, a soft assertion reports how to change these values in order to trigger the instability. We then use the output of soft assertions as signals to effectively mutate inputs to trigger numerical instability in ML applications. In the evaluation, we used the GRIST benchmark, a total of 79 programs, as well as 15 real-world ML applications from GitHub. We compared our tool with 5 state-of-the-art (SOTA) fuzzers. We found all the GRIST bugs and outperformed the baselines. We found 13 numerical bugs in real-world code, one of which had already been confirmed by the GitHub developers. While the baselines mostly found the bugs that report {\tt NaN} and {\tt INF}, our tool \tool found numerical bugs with incorrect output. We showed one case where the \textit{Tumor Detection Model}, trained on Brain MRI images, should have predicted "tumor", but instead, it incorrectly predicted "no tumor" due to the numerical bugs. Our replication package is located at \url{https://figshare.com/s/6528d21ccd28bea94c32}.

\end{abstract}

%\begin{keywords}
%Numerical Bugs, Fuzzing, Assertion, Machine Learning, Software Engineering.
%\end{keywords}

\maketitle
\section{Introduction}
\mycomment{
{\bf Motivation}:
A numerical bug in deep learning applications can waste training time and resources. When low accuracy is reported, it is hard to debug. We do not know whether it is due to insufficient data, hard problem, bad model architecture, incorrect use the APIs and their parameters, or numerical bugs. Numerical bugs often are silent or can be propagated to report NAN and INF. It is hard to trace back to the root cause during debugging, typically in some unstable function. Thus, we have the motivation of generating the inputs and finding the numerical bugs with assertions.

\wei{explain what is unstable functions here}

\miryung{I think we may want to clarify how neural assertions can be inferred or specified or how much training data is needed (or could be generated)}
\miryung{regarding the second contribution, we may want to clarify how it is different from deepinstability work?}
Contributions:
\begin{enumerate}
\item novel ideas of using soft assertion to capture fault conditions that cannot be easily expressed using traditional symbolic assertions
\item an approach to automatically infer soft assertions for the use of numerical unstable functions
\item a database of numerical unstable functions
\item a framework and tool of fuzzing with soft assertion to detect numerical bugs in DL applications
\item a comprehensive evaluation
\end{enumerate}

}

Machine learning (ML) applications are becoming essential in a wide range of domains~\cite{hawkins2021guidance,giger2018machine, gomez2015netflix}. ML code heavily relies on floating-point computation and often involves very large/small numbers, posing a significant challenge of {\it numerical stability} of computation. We say a function is {\it numerical unstable} when a small change in the input can lead to large, and often unexpected differences, in the output~\cite{burden1997numerical,di2017comprehensive, kloberdanz2022deepstability}. For example, {\tt division} is an unstable function---we can divide a value close to zero successfully; however, when the value becomes a bit smaller and is truncated to zero, the division reports an {\tt INF} (when the divisor is zero) or {\tt NaN} (when the dividend and divisor are both zeros). When such an error occurs, we say that {\it numerical instability} or {\it numerical bug} is triggered. 

ML applications perform machine learning tasks typically using APIs from frameworks such as PyTorch or TensorFlow. The application code as well as the API implementations can contain unstable functions. When unstable functions are not properly guarded, there can exist a feasible input to trigger numerical instability in these unstable functions, which then causes the applications to crash, generate low accuracy and incorrect predictions, and/or stop learning but continue wasting computational resources during training~\cite{jia2020empirical}. The numerical bugs have been rated as {\it high priority} by professional ML developers~\cite{kloberdanz2022deepstability}.

%In our study, we reproduce\wei{For example, recent research~\cite{} reported that{\tt yolo}~\cite{gitrepo}, a popular ML application, produces incorrect image segmentations due to numerical bugs.}

%We say that the code contains a numerical bug when the unstable functions are used and there exists a feasible input that can trigger the numerical instability. When numerical bugs occur in ML applications, the applications can crash, stop learning but continue wasting computation resources during training, and/or generate low accuracy and incorrect predictions~\cite{jia2020empirical}.

%For example, we can divide a value close to zero successfully; however, when the value becomes a bit smaller and is truncated to zero, the output becomes a {\tt INF} or {\tt NaN} (when the dividend is small). The functions can manifest such behaviors under certain inputs are called {\it unstable functions}. \miryung{add one example of unstable function and how its failure can trigger unexpected output? make it concrete?} 
Numerical bugs are challenging because (1) unstable functions commonly exist in ML libraries and applications~\cite{kloberdanz2022deepstability} but ML developers may not know them; (2) even for ML experts, it is very hard to understand numerical properties of complex functions to add a proper assertion before invoking unstable functions; (3) in many cases, only 
special crafted inputs can trigger the bugs while the inputs can be complex tensors which traditional static analysis and testing tools cannot handle; and (4) numerical bugs do not always have a failure symptom of {\tt NaN}/{\tt INF}, and it can lead to an incorrect output that cannot be detected without specially crafted {\it oracles} (the oracle  provides a ground truth output from a stable implementation).%, comparing which, we can demonstrate that a numerical bug is triggered in the unstable function).

In this paper, to address these challenges, we constructed a database of 61 unstable functions that can be referenced by ML developers. We analyzed the numerical properties and designed 6 types of oracles for the 61 unstable functions. We developed an automatic approach to encode safety/error conditions of invoking unstable functions via machine learning, using which we designed a novel fuzzing technique that successfully generates inputs to trigger numerical instability in ML applications.

Our database of unstable functions leveraged prior literature of DeepStability~\cite{kloberdanz2022deepstability}, where the authors identified unstable functions for PyTorch and TensorFlow libraries. We expanded the set of unstable functions and also provided more detailed information for each unstable function, such as oracles and potential input conditions that trigger the numerical bugs. There are GRIST~\cite{yan2021exposing}  and RANUM~\cite{li2023reliability} which can automatically detect numerical bugs in ML applications. Their work used manually encoded error conditions and thus only handled a few unstable functions. In addition, they only detected numerical bugs that led to {\tt NaN} and {\tt INF}. Our work aims to handle much more unstable functions via automatically encoded error conditions. We design test oracles that enable to find numerical bugs with failure conditions beyond {\tt NaN} and {\tt INF}. Other related work also includes PositDebug~\cite{chowdhary2020debugging,chowdhary2021parallel}. It detects numerical bugs by using an alternative floating-point representation, {\it posit}, offering higher precision through shadow execution with high-precision values.  DEBAR~\cite{zhang2020detecting} uses abstract interpretation to analyze whether the value of a variable violates the expected range. They covered 8  unstable functions but faced the challenge of false positives. 

Specifically, in this paper,  we developed \tool to automatically detect numerical instabilities in ML applications. Our key novel idea is the {\it soft assertion}, aimed at using machine learning models to capture safety/error conditions of unstable functions. It addressed a critical challenge that developers cannot easily (in fact, sometimes impossibly) write symbolic assertions or checks in the code to assure the numerical stability; such conditions require an in-depth understanding of numerical properties of library calls. In this paper, we first generate a soft assertion by testing an unstable function alone, such as a framework API. We then apply the soft assertions to guide fuzzing for any ML applications that use the unstable functions. The fuzzer will generate inputs  at the ML application level to show how the numerical instability is triggered in ML applications, if any. 
Compared to existing fuzzers which typically used random input generation~\cite{pyfuzz2022, hypothesisDocs} or code-coverage guided search~\cite{atheris2023}, \tool mutate inputs based on the error conditions provided by soft assertions. We have shown, in our evaluation, that such assertion guided fuzzing is much more effective to trigger bugs than existing fuzzers. %in real-world code.

%he soft assertion, once generated, can be used for any ML applications that invoke the unstable function. 
%we used soft assertions to guide fuzzing; that is, we first test an unstable function {\it only} to collect data to generate a soft assertion, and then we use it to generate inputs

%\miryung{avoid term test, instead execution profile? since we also use train vs. test in the ML cross evaluation}
Our techniques consist of two main components. The first component is the generation of soft assertions.
For each unstable function we collected in the database, we automatically generated inputs and performed unit testing to collect successful and failure-inducing inputs for the unstable function.   This execution profile data are then converted to the labeled data to train an ML classifier (soft assertion), where the input is the matrices input to the unstable function, and the label is a {\it soft assertion signal} that instructs how we should transform a passing input to trigger the numerical bugs. As an initial exploration, we used a set of predefined transformations, including  {\it increase}, {\it decrease} and {\it no-change}, meaning that we should increase/decrease/not change the current value in order to trigger the numerical bug in the unstable function. We plan to expand to other labels in the future work.

Our second component is a fuzzer that uses soft assertions as guidance. Given an ML application, we first scan the code to identify the unstable functions. During fuzzing, when an execution reaches an unstable function, we  activate its soft assertion to evaluate the current input values. If we obtain a {\it no-change}, we will execute the values to confirm that numerical instability is triggered; otherwise, we will use {\it auto-differentiation}~\cite{paszke2017automatic} to propagate the soft assertion signal (the transformation advice) to the entry of the application code to inform how the current input should be mutated.  {While coverage-guided fuzzing simply mutates inputs that expand coverage, this soft assertion provides concrete guidance on which direction to mutate the input values.} This iteration terminates until we trigger the numerical instability or until a timeout is reached.

%when we receive a {\it signal} to guide how to mutate the value to trigger the failure in the unstable function, we will propagate such signal to the input via an {\it auto-differentiation}~\cite{paszke2017automatic} component to guide the transformation for the input.

We implemented \tool for Python programs that use PyTorch or TensorFlow frameworks. In evaluation, we constructed a benchmark consisting of 79 programs from GRIST~\cite{yan2021exposing} where we know the numerical bugs, and 15 real-world applications from GitHub with a 10 star rating, where we do not know the numerical bugs. We detected a total of 92 numerical bugs including all the 79 bugs in the GRIST benchmark and 13 unknown bugs in real-world code. Our tool significantly outperformed the 5 state-of-the-art Python fuzzers: (1) we found all the GRIST bugs, but more efficiently,  (2) we found 13 real-world bugs, 1 of which already confirmed by the GitHub developers while the baselines hardly found any, (3) we found numerical bugs that lead to the incorrect output and incorrect ML predictions, while the baselines mostly handled the bugs that lead to {\tt NaN/INF}. We presented a case study where a tumor detection model, TumorScope~\cite{ID_15}, trained on Brain MRI images incorrectly predicted "no tumor" after triggering a numerical instability. 

In summary, our paper made the following contributions:
\vspace{-.1cm}
\begin{enumerate}
    \item We built a database, consisting of 61 numerically unstable functions---56 are collected from the prior literature~\cite{kloberdanz2022deepstability} and 5 are discovered in our research. Compared to~\cite{kloberdanz2022deepstability}, our database included further details for each unstable function, such as oracles and potential input conditions that trigger the numerical bugs;
    \item We designed 6 types of oracles that enable us to find failures beyond {\tt NaN} and {\tt INF};
    \item We proposed a novel idea of {\it soft assertion}, where we use ML models to automatically capture safety/error conditions of a numerical unstable function; our experience is that manually specifying symbolic assertions in these cases is almost impossible;
    \item We developed an approach to automatically infer soft assertions by testing numerically unstable functions against our oracles;
    \item We developed \tool that used the feedback of soft assertions to automatically generate inputs 
    to trigger numerical bugs in ML applications. The user can directly use our tool to scan their code for numerical instability supported by soft assertions we generated so far. The user can also expand to more unstable functions by generating soft assertions using our framework and our approaches, and then plug them into our fuzzer; 
    \item We performed a comprehensive evaluation on 94 ML applications and 5 SOTA baseline fuzzers, and demonstrated that our tool can find numerical instabilities in real-world ML code where existing tools cannot. We also show that  besides {\tt NaN/INF}, numerical bugs can lead to incorrect model predictions (we have not seen such examples in prior literature).
    \newline
\end{enumerate}

\section{A Motivating Example} 
\vspace{-10pt}

\begin{figure}[!ht]
\centering
\includegraphics[width=\linewidth]{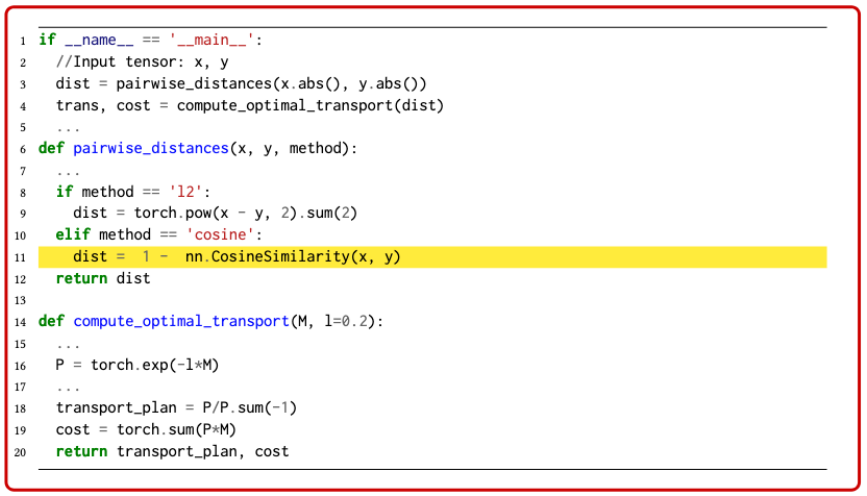}
\caption{Unstable {\tt CosineSimilarity} leads to incorrect results in PyTorch}
\label{fig:example_2}
\end{figure}

%added from rebuttal 

%overview
In Figure~\ref{fig:example_2}, we show a numerical bug in a snippet of machine learning code that triggers the unstable function, {\tt CosineSimilarity}, implemented in PyTorch~{2.2.1}. The code takes two tensor inputs that represent a source and a target; it aims to find an efficient strategy to allocate resources from the source to the target. The numerical bug has led to an incorrect output of this program and incorrectly computed the cost of resources. For example, given the input: 
\[ x = \begin{bmatrix}
2606.66824394 & 2477.72226966 & 3251.84008903
    \end{bmatrix}\] \[y = \begin{bmatrix}
1.4682216\times10^{-9}&1.825367\times10^{-8} &8.3524\times10^{-9}
\end{bmatrix}\] the correct output {\tt cost} should be 0.08963637, but the actual output {\tt cost} is 0.1601. We used an oracle to determine the correct output. This oracle compared the PyTorch implementation with a more stable implementation of cosine similarity in NumPy. See Section~\ref{oracle} for details.

%The correct {\tt cost} value was determined using the {\tt Consistency Check Across Frameworks} oracle (Table~\ref{tab:sample-table}), where we compared the PyTorch implementation against a reference implementation of cosine similarity that we created using NumPy.
% The correct output {\tt cost} value of 0.08963637 is determined using oracle named Consistency check across frameworks listed in table~\ref{tab:sample-table}, where we compare the PyTorch implementation against a reference implementation of cosine similarity that we created using NumPy.

\noindent{\bf Analysis of the numerical bug:} The two inputs, {\tt x} and {\tt y}, represent a set of characteristics for the source and the target respectively. The {\tt pairwise\_distances} function at line~3 first evaluates the discrepancies between the two inputs. At line~4, this distance input is passed to the {\tt compute\_optimal\_transport} function. This function (details at line 14) takes the input of discrepancies and report {\tt P} at line~16; {\tt P} represents how to redistribute resources from the source to the target in a manner that can minimize the total cost.  Finally, {\tt transport\_plan} is generated after normalizing {\tt P} at line 18, and {\tt cost} is calculated at line 19.

The root cause of this error is that the input triggered the instability of {\tt CosineSimilarity}. The cosine similarity between two non-zero vectors can be calculated using the formula for the Euclidean dot product:
\[cosine\_similarity(x,y) = \frac{x.y}{||x||*||y||} \] When the {\it norm} ($||\cdot||$) of any input is less than $10^{-8}$, the PyTorch implementation will assign the norm value to $10^{-8}$. In this example, the norm of $||y||$ should be $9.2263\times10^{-9}$, but $||y||$ is adjusted to $10^{-8}$. As a result, the cosine similarity results in 0.8399 instead of 0.91036362. This deviated value of {\tt dist} at line~11 is then propagated in the code and lead to the incorrect output of {\tt cost} at line~20. In the past, other researchers~\cite{kloberdanz2022deepstability} have also reported the instability of cosine similarity, where the bug produced an output larger than 1, outside its valid range of~[-1,1].

%This error is then propagated in the code and lead to the incorrect output of {\tt cost}. Other researchers~\cite{kloberdanz2022deepstability} have also reported the instability of cosine similarity but triggered in a different way. There, the bug produced an output larger than 1. This value is outside its valid range of~[-1,1].

\noindent{\bf Challenges of avoiding and detecting such bugs:} This type of numerical bug is hard to avoid or detect; unlike a divide-by-zero error where a developer may know to add a check or assertion before the division, stating that {\it the divider should not be a zero}, here, it is not easy to write a check or assertion before or after {\tt CosineSimilarity} to detect the error.  This is because some ML functions are mathematically complex, and it is challenging to understand when the function could become unstable~\cite{kloberdanz2022deepstability}.  Existing work on mining symbolic preconditions, e.g., Daikon~\cite{Daikon}, cannot work here either (see Section~\ref{threat})---- in ML applications, the stability conditions involve high-dimensional matrices and vectors, e.g., it is hard to express ``the determinant of a high-dimensional matrix (of unknown size) is zero'' in a symbolic condition. 

%the authors present a similar view with more detailed examples.

%Existing symbolic preconditions are not easily applicable to machine learning (ML) functions because (1) traditional approaches face challenges in handling preconditions related to high-dimensional matrices and vectors; for example, it is hard to express ``the determinant of a high-dimensional matrix (of unknown size) is zero'' in a symbolic condition, and (2) some ML functions can be mathematically complex, and thus it is even hard for domain experts to understand when the function could become unstable and to be able to manually specify the assertion as symbolic conditions. For example, \texttt{CosineSimilarity} is an unstable function, and we are not able to manually specify the stable condition as logical constraints; however, when we trained a “soft assertion,” it detected the instability. \cite{kloberdanz2022deepstability} presents a similar view with more detailed examples.

\noindent{\bf Our novel idea to address the challenges:} Although it is hard to express the condition of numerical stability as symbolic constraints, we trained an ML model, namely {\it soft assertion}, to encode such conditions and use it as an assertion to detect numerical instability and give feedback to guide fuzzers.  Soft assertions are inserted at the call to a numerical unstable function, which can be either an API to the library or a user-defined function. Soft assertion is an automatically generated ML model that encodes the error-triggering conditions for an unstable function. The training dataset consists of failure-inducing and successful inputs automatically obtained via performing unit tests of an unstable function. For each unstable function like {\tt CosineSimilarity}, the training is only performed once to generate its soft assertion, and once generated, the soft assertions can then be repeatedly used in any ML applications that use the function. Like typical program assertions, soft assertions may be used for runtime monitoring and/or serving oracles for testing; in this paper, we focus on exploring its applications for detecting numerical instability and effectively guiding fuzzers to trigger numerical bugs in ML applications.

\section{Soft Assertion}

\subsection{What is a Soft Assertion?}~\label{sa}
%\wei{this section is done by wei}
%compare neural assertion with symbolic assertion to explain its advantage

Soft assertion is similar to a typical symbolic assertion in that we insert it at a check point of a program to detect errors. The term 'soft' reflects the fact that these assertions are not precise symbolic conditions; instead, it encodes (approximate) error conditions in a machine learning model when symbolic conditions cannot be easily specified.  We view ``soft assertion'' as one kind of ``assertion'' because it takes program values at the assertion point and reports whether the error condition is violated. When the assertion is not triggered, it provides advice on whether the current values should be {\it increased} or {\it decreased} in order to trigger the assertion.  Such functionalities are similar to symbolic assertions, which can also be used to trigger failure conditions and guide testing.

In this paper, we apply soft assertions to find numerical bugs, and therefore, soft assertions will be inserted before unstable APIs (the unstable functions implemented in the library) or calls to the user-defined unstable functions in an ML application. We have collected a list of unstable functions through our research and also from prior literature~\cite{kloberdanz2022deepstability}; the users can also use our approach to generate soft assertions for their own defined unstable functions.

\vspace{0.1cm}
\noindent{\bf Definition}: A {\it Soft Assertion (SA)} at a program point $p$ is a machine learning model that takes the program values, denoted as $X$, at $p$, and reports how to transform $X$ to trigger the error condition at $p$. Mathematically, we denote 
\begin{equation}~\label{eq1}
SA(X) = 
\begin{cases}
increase  \\
decrease  \\
no\; change
\end{cases}
\end{equation}
where $X$ consists of program values at $p$.  The output $SA(X)$ reports three types of actions: {\it increase} (increasing the values in $X$), {\it decrease} (decreasing the values in $X$) and {\it no change} (the error triggered and the failure-inducing values found). We call the output of the soft assertion, $SA(X)$, {\it Soft Assertion (SA) Signals}. Our current design of formula (1) tailors the applications of guiding test input generations in fuzzers. In the future, we plan to expand soft assertions to more classes and explore applications beyond fuzzing.

We designed the concept of soft assertion by referencing the roles of a traditional symbolic assertion. An assertion can test if the error condition is satisfied, which corresponds to the {\it no change} signal in formula~(1). Traditional symbolic assertions can also be incorporated by test input generators, such as symbolic execution, as constraints to generate test inputs. Considering that  writing symbolic assertions for high dimensional tensor values and complex numerical functions are extremely challenging, and symbolic execution cannot be easily applied, in this paper, we apply fuzzing for test input generation and explore alternative signals (instead of concrete symbolic conditions) to guide test input generation. Specifically, we used a simple classification formulation and designed {\it increase} and {\it decrease} two SA signals, as show in formula (1), to provide the directions of input mutation in fuzzing, as we are fully aware that it is very difficult to require an ML model to generate complex symbolic conditions~\cite{watson2020learning, white2020reassert}. Later, we show in Section~\ref{sec:results}, even with our simple formulation,  SA signals can be effectively used to mutate program inputs to trigger the bugs.

\subsection{How to Generate a Soft Assertion}
%\wei{done by Wei}
In this paper, we developed an approach to automatically generate soft assertions. It only requires the unstable function as input, and the generation is independent of the ML applications that call the unstable functions. The soft assertion of an unstable function only needs to be trained once, and then it can be inserted in different ML applications where the unstable function is called.

\begin{figure}[htbp] % htbp is a placement specifier
  \centering
  \includegraphics[width=\textwidth]{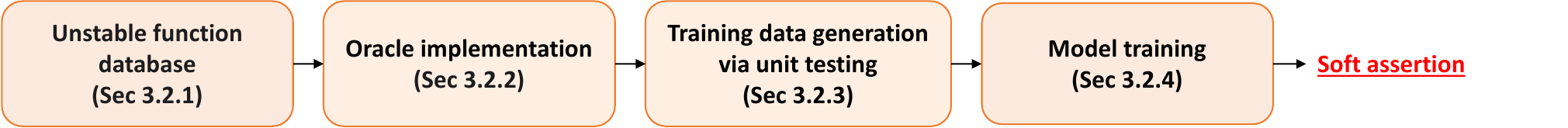}
    \caption{An Overview of Soft Assertion Generation}
    \label{fig:nagen}
\end{figure}

In Figure~\ref{fig:nagen}, we show our approach, which consists of four steps. First, we collected a list of frequently used, known unstable functions (Section~\ref{database}). To build a soft assertion, we developed the oracles to determine whether the numerical bugs were triggered at the end of the numerical function (Section~\ref{oracle}). Using the oracles, we ran unit tests on individual unstable functions to record the failure-inducing and successful values at the entry of each unstable function. In the next step, we designed a test input generation technique to efficiently generate the balanced training dataset consisting of successful and failure-inducing inputs (Section~\ref{train}). Finally, we trained a ML model for each unstable function that can predict the actions of $SA(X)$ in formula~\ref{eq1}. In the following sections, we will describe each step in detail.

%To generate an assertion at a point of these calls, we train a neural networking via unit testing of the unstable functions. Specifically, we randomly generate the test inputs for each unstable function, and classify these inputs as failure-inducing or successful. To do so, we design and implement different types of oracles that can report numerical bugs. We then use these data and train a neural network that performs the classification tasks of $NA(X)$ in Eq~\ref{eq1}. In the following sections, we will describe each step in details.

%an unstable function database. We studied a set of numerically unstable functions from the literature ~\cite{yan2021exposing,zhang2020detecting,kloberdanz2022deepstability} and developed the oracles of stable implementations. Each unstable function will have a neural assertion generated. The training data of neural assertions are generated by doing unit testing on individual unstable functions. We identified and implemented the oracles for a representative set of unstable functions (Section~\ref{oracle}) so that we can obtain successful and failure-inducing inputs. In Section~\ref{train}, we provide the details on training data generation; as shown in Figure~\ref{fig:nagen}, it is an iteration loop of generating test inputs (neural assertion inputs) and curating the transformations that potentially lead to the failure (neural assertion labels). In Section~\ref{model}, we explain our model training techniques.

\subsubsection{\textbf{Building a Database of Unstable Functions}}~\label{database}
%\wei{done by wei} 
Machine learning applications frequently invoke APIs from the libraries and frameworks. Numerical errors can be largely managed if we can identify frequently used unstable APIs and ensure that when they are called in applications, their instability (error conditions) cannot be triggered.

%However, it is hard to add a checker to prevent the error conditions being triggered. As we mentioned above, the conditions of triggering the instability are hard to specify when the inputs are vectors and tensors.
We studied the prior literature~\cite{kloberdanz2022deepstability}  and collected a total of 56 unstable functions from the PyTorch and TensorFlow libraries. During the study, we found additional 5 unstable functions, and our database thus includes a total of 61 unstable functions. These functions are critical to ensure numerical stability and performance in modern ML applications. The functions consist of activation functions, derivative computation and {\it hyperbolic} functions (see our replication package for the entire list of functions). For each unstable function, we attempted to manually identify input conditions that trigger numerical bugs and succeeded for 11 functions.  We have tried the binary search on the input range and used extremely large and small boundary values as well as diverse input patterns of a matrix, e.g., sparse matrices and ill-conditioned matrices.  It is consistent with what we mentioned above that it is hard to create a symbolic assertion or insert a checker in the application code to avoid the numerical errors of the unstable functions being triggered. While the prior literature listed many of those unstable functions~\cite{kloberdanz2022deepstability} and explained why they are unstable for some functions, our work analyzed all 61 unstable functions in our database, and identified test oracles for these functions.
% Using binary search, extreme values (both very large and small), and diverse input patterns such as sparse and ill-conditioned matrices, we systematically explored instability conditions. 
% Our findings reinforce the challenge of constructing symbolic assertions in application code to prevent numerical errors in unstable functions. The 11 identified functions exhibited instability primarily through the occurrence of NaNs/Infs in outputs and mismatches with reference solutions. 

%As we mentioned above, the conditions of triggering the instability are hard to specify when the inputs are vectors and tensors.

\subsubsection{\textbf{Designing and Implementing Oracles}}~\label{oracle} %\wei{done by wei}
To generate soft assertions, we need to identify the oracles that can distinguish the correct and incorrect output of unstable functions. To design the oracles for numerical bugs, we first conducted a thorough analysis of the mathematical properties and potential sources of instability of the 56 unstable functions in our database, using the prior literature~\cite{yan2021exposing,zhang2020detecting,kloberdanz2022deepstability}  as a reference. This analysis allowed us to identify specific types of numerical errors and failure symptoms these functions might produce such as undefined results, out-of-range results, precision loss/incorrect results, algorithm instability, and inconsistencies across implementations that use different libraries. {For each error category, we formulated a type of oracle designed to detect that specific class of error.} Our analysis resulted in {6 types of} oracles, summarized in Table~\ref{tab:sample-table}. In the following, we provide the detailed explanations:

\begin{table}[]
    \centering
    \caption{Constructing Oracles to Test Unstable Functions}
    \label{tab:sample-table} 
    %\begin{tabular}{lccc}
    \begin{tabularx}{\textwidth} {p{4cm} p{4cm} X }
    \toprule
    \textbf{Oracle} & \textbf{Example Unstable Function} & \textbf{Math Properties of Unstable function}\\
        \midrule
        NaN/INF detection & Softmax, L2 Norm, Division  & The function has a division operator; the divisor can be small or 0\\ \midrule
        \addlinespace[-1pt]
        Out-of-range check & Sin(x), Cos(x), Inverse tangent, Hyperbolic tangent & The function has known bounded outputs. \\ \midrule
        \addlinespace[-1pt]
        Math formula rewriting & x-(max + log(y))
        
        x-y $\times$ log(x)
        
        & 
        The function has a subtraction operator, and the operands have very similar values; or it contains a log operation that can take very large/small input. 
       % \textcolor{orange}{In some cases, replacing division with multiplication or rearranging terms can improve numerical stability.}
       % operation or log operation. The values of the operands can be very similar. Log operation may lose precision for very large/small input
        % Subtraction of nearly similar numbers.  Also may loss precision for very large or small value of x and y
        \\ \midrule
        \addlinespace[-1pt]
        Stable algorithm implementation& Matrix Inverse, Matrix determinant, Eigen value decomposition
        & The function uses complex algorithms that are highly sensitive to ill-conditioned matrices or matrices that are nearly singular.
        \\ \midrule
        \addlinespace[-1pt]
        Consistency check across frameworks& Cosine similarity, Linear solvers, Eigen value calculation, SVD, Random number generator & Some functions are known for their sensitivity to numerical stability, so we need to check across different library implementations
        % For example, cosine similarity, can produce different results across libraries (e.g., NumPy vs PyTorch) due to differences in precision, optimization, and algorithms
        \\ \midrule
        \addlinespace[-1pt]
        Increase width & Remainder, Fast Fourier transform, polynomial roots, PCA & The function uses operations that can lead to precision loss when the values of operators have different magnitudes (small/large input).     
        %The function has a modulus operator. The dividend can be large, or the divisor can be small which may lead to incorrect results due to differences in numerical precision.
        % Sensitive to floating-point rounding for large x or small y
        \\ \addlinespace[-1pt]
    \bottomrule
    \end{tabularx}
\end{table}

%based on the prior literature~\cite{yan2021exposing,zhang2020detecting,kloberdanz2022deepstability} as well as our domain knowledge of numerical stability. We summarized them in Table~\ref{tab:sample-table}.
%\textcolor{orange}{Review's concern: Rationality discussion of using oracles: \\ Our approach to developing oracles for numerical instability detection began with a manual inspection phase, which laid the groundwork for a more automated, systematic method. Initially, we  categorized unstable functions documented in prior literature~\cite{yan2021exposing,zhang2020detecting,kloberdanz2022deepstability} and conducted a thorough analysis of their mathematical properties and potential sources of instability. This analysis allowed us to identify specific types of numerical errors these functions might produce such as undefined results, out-of-range results, precision loss/incorrect result, algorithm instability, and inconsistencies across implementations. }
%\textcolor{orange}{For each error category, we formulated an oracle type designed to detect that specific class of error.}The first two types of oracles define erroneous outputs when instability is triggered:

\noindent{\textbf{(1) NaN/INF detection}:} Some unstable functions fail with a {\tt NaN} or {\tt INF}, indicating undefined or infinite results. We invoke {\tt isNaN and \tt isInf} (implemented in the ML libraries) to test the return value of the unstable function to determine if the unit test result leads to a NAN/INF. For example, a division of a very small number can lead to a NAN/INF. Here, the division is an unstable function, and the NAN/INF check is used as an oracle. We also identified {\tt SoftMax} and {\tt L2Norm} as this category based on our analysis of these functions. 

\noindent{\textbf{(2) Output-of-range check}:} For some unstable functions, we can identify a valid range. For example, {\tt sin(x)} and {\tt cos(x)} have a valid range of [-1,1], any deviation from the values will be flagged as undesirable output. Sometimes, even if the return value is within the valid range, numerical instability can still occur as shown in the example of Figure~\ref{fig:example_2}, so we will consider all types of oracles that apply for an unstable function.

%Often, numerical bugs can lead to incorrect results and precision loss. To detect them, we need to construct a (relatively) stable version of the unstable function to compare their results.

\noindent{\textbf{(3) Math formula rewriting}:} Some math formulas are unstable, but we can create more stable implementations through rewriting~\cite{kloberdanz2022deepstability}. For example, we can rewrite $x/(sqrt(x)*sqrt(x))$ to $x/sqrt(x*x)$ to avoid precision loss, and rewrite
$x-(max+log(y))$ to $x-max-log(y)$ to avoid the loss of significant digits. We implemented a stable version of the function using math formula rewriting as an oracle. We compare the output from the current unstable equation with the output from the oracle.

\noindent{\textbf{(4) Stable algorithm implementation:}}
Some algorithms may produce mathematically accurate results but lead to numerical issues for some input. For example, the algorithm used to calculate the direct inverse of matrix exhibit instability. If the input matrix is ill-conditioned such as the matrix is close to the singular matrix, the result can be incorrect. Here, the {\tt inverse} operation is the unstable function, and we use a more stable implementation as an oracle~\cite{kloberdanz2022deepstability} for the inverse calculation, namely {\it Cholesky inverse}.

\noindent{\textbf{(5) Consistency check across frameworks}:}
Different library frameworks sometimes have different implementations for the same math functions, resulting in different numerical stability properties for the functions. For example, we found that some inputs to the {\tt cosine similarity} function can lead to very different outputs across NumPy, SciPy, and PyTorch. Compared to our manually calculated results, we found that NumPy implementations are the most accurate and thus we used it as oracles. In a pilot study, we generated two 1024 vectors of randomly generated values ranging from $10^{10}$ and $10^{-10}$, our manually calculated values and NumPy both report 0.753396; however, PyTorch reports 0.137953. When implementing this oracle, we are aware that we cannot always exactly match the two floating point numbers even if they are both correct; therefore, we consider a match when the return values of two implementations have a difference within a configurable threshold (we set 0.000001 in our experiments following the literature~\cite{dawson2008comparing}).

%\textcolor{orange}{Dr. Kim's Concern:However, it is difficult to check the floating point values for either consistency check or precision loss or incorrect result.While comparing we want to know if the results are close to each other. Therefore, we consider an method of relative error with epsilon. We are considering 99.999\% of accuracy , therefore our threshold is 0.000001} \cite{dawson2008comparing}

%, e.g., when the elements in one vector/matrix have very large differences compared to the ones in another vector/matrix,

%Same operation might result in different outcomes across various libraries due to implementation and floating-point precision differences. For instance, calculating cosine similarity across NumPy, SciPy, and PyTorch can result in varying outcomes, especially with vectors/matrices spanning a broad range of magnitudes. In such cases, NumPy or frameworks operate with higher default precision is considered as an oracle.
\noindent{\textbf{(6) Increase the width of floating-point representation}:} For the cases where none of the above oracles are applicable, we consider to increase the width of the floating point numbers.  For instance, a remainder calculation with 32-bit floats might not accurately represent certain decimal numbers. {As an example, {\tt 1933053808.0\%53} leads to 35 instead of the correct value 19 when calculated in the 32-bit precision. Thus, in the oracle, we implement the {\tt remainder} operation using the 64-bit precision to represent its operands.}

\newcommand{\specialcell}[2][c]{%
  \begin{tabular}[#1]{@{}c@{}}#2\end{tabular}}

\newcolumntype{P}[1]{>{\raggedright\arraybackslash}p{#1}}
\begin{table*}[!ht]
\centering
\caption{Numerical Stability Analysis of Newly Identified Functions}
\label{tab:new_functions}
\small
\begin{tabular}{P{2.5cm}P{3cm}P{4cm}P{2.5cm}}
\toprule
\textbf{Function} & \textbf{Safe Condition} & \textbf{Source of Instability} & \textbf{Oracle} \\
\midrule

Linear Solver \newline $(Ax = b)$ & 
Unknown & 
(1) Large condition number \newline
(2) Near-singular matrix ($det(A) \approx 0$) 
% $\bullet$ Zero/near-zero pivots 
& 
Increased width  \\
\midrule

SVD \newline (Singular value decomposition) & 
Unknown & 
(1) Near-zero singular values \newline
(2) Ill-conditioned input matrix \newline
(3) Loss of orthogonality & 
Consistency check across frameworks \\
\midrule
Eig \newline
(Eigenvalue Computation) 
& 
Unknown & 
(1) Ill-conditioned eigenvectors \newline (Repeated eigenvalues, large eigenvalue ratio) & 
Increased width\\
\midrule

SELU/ELU 
% \newline (Scaled Exponential Linear Unit) / (Exponential Linear Unit)
&
% $x_i \leq 88.722839$ (32-bit)
% % 709.78271289338 - actual value
%  \newline $x_i \leq709.78271$ (64-bit)
% \newline [SELU: $\alpha $=1.6732632, $\lambda=$1.0507009\newline ELU: $\alpha $ = 1]
-$103.9720 \leq x_i \leq 3.40e38 $ 
 % \newline 
 % $x_i \geq$-745.97 
& 
(1) Certain negative inputs in tensor (vanishing gradient)\newline
(2) Large positive inputs in tensor (overflow)& 
NaN/INF detection \\
\midrule

GlorotUniform \newline
(Weight Initialization) & 
Unknown & 
(1) Improper scaling \newline
(2) Poor weight distribution & 
Stable algorithm implementation \\
\bottomrule
\end{tabular}
\end{table*}

 Table ~\ref{tab:new_functions} provides an overview of five functions identified in our research that were not reported in prior works. We list the rest of unstable functions in our replication package, including the oracle type(s) for each unstable function. We implemented oracles for 25 unstable functions. In the future, if the user wants to extend our framework to support a new unstable function, they can determine which one(s) of the 6 oracles matches the potential numerical errors the unstable functions will produce. It will be useful to analyze the mathematical properties of the function, for example, the domain and range of the function, or the potential sources of instability such as subtraction of similar numbers, and division by small values). See the third column in Table~\ref{tab:sample-table} for the mathematical properties we considered when determining oracles.

\subsubsection{\textbf{Performing Unit Testing on Unstable Functions to Create Training Data}}~\label{train}
%\wei{done by wei}
Our goal is to train an ML model that approximates formula (1), defined in Section~\ref{sa}.  Specifically, the model input $X$ is the test input of the unstable function, and the model output is the transformation we should use to mutate $X$ to trigger the numerical instability in the unstable function.
In Table~\ref{tab:trainingdata}, we used examples to explain how we produce the training data using unit testing of unstable functions.

% Our approach is to automatically generate test inputs and perform unit testing for the unstable function to obtain the training dataset.

\begin{table}[ht]
\centering
\caption{Our approach of generating training data}
\label{tab:trainingdata}
\resizebox{0.7\textwidth}{!}{
\begin{tabular}{ccccc}
\hline
\toprule
\textbf{Base} & \textbf{Mutation} &  \textbf{Current} & \textbf{Output Change} & \textbf{Training Data}  \\ \addlinespace[-1pt]
\midrule
\multirow{2}{*}{10} & \multirow{2}{*}{increase} & \multirow{2}{*}{30} & \multirow{2}{*}{fail $\rightarrow$ success} & (30, decrease) \\  \addlinespace[-1pt]
\cline{5-5}
 &  &  &  & (10, no change) \\ \addlinespace[-1pt]
\hline
\multirow{2}{*}{-1} & \multirow{2}{*}{increase} & \multirow{2}{*}{5} & \multirow{2}{*}{success $\rightarrow$ fail} & (-1, increase) \\
\cline{5-5}
 &  &  &  & (5, no change) \\ \addlinespace[-1pt]
\hline
\end{tabular}
}
\end{table}

Suppose we randomly generate a test input 10 (note that in real applications, it will be a matrix), namely the {\it base} input, as shown in the first row of the table. When running this input, the numerical instability of the unstable function is triggered. We then apply a mutation to this input(without the loss of generality, let's say it becomes 30). This input leads to a success in testing. Through these two tests, we obtain two training data points: (1) (30, decrease), meaning that given a successful input 30, if we decrease it (e.g., to 10), the unstable function likely fails; and (2) (10, no change), meaning that given 10, the unstable function will fail and no change is needed. Similarly, in the second example, shown in the next row, our base input is -1. Suppose it leads to a successful execution. After mutation, when we increase its value to 5,  the test fails. We obtain the training data points (-1, increase) and (5, no change).

We aim to generate a balanced training data set, i.e., obtaining the similar numbers of labels of "increase", "decrease" and "no change", and we also hope the input set can cover a diverse set of values. We developed two key components (1) the approach of mutating inputs, and (2) the approach of generating initial base inputs used for mutations:

\noindent{\bf (1) Mutation approaches:} The goal of the input mutation is to obtain a new input whose test outcome would differ from the base input, i.e., one fails and the other passes during unit testing of an unstable function. The transformation between the two inputs can then be used to generate labels as indicated in Table~\ref{tab:trainingdata}.

Triggering numerical stability is very challenging, and our mutation supports a set of configurable methods to do the step-wise increases/decreases, starting from the base input. Specifically, {we implemented the {\it exponential} ($step = e^{rate}$), {\it random} ($step = r*rate$), and {\it sinusoidal} ($step = sin(rate)$) three methods for choosing different speeds of changes during mutation.} {\it rate} here is a configurable parameter. To further increase our chance of obtaining diverse data, we mutate from failure-inducing inputs to success inputs (shown in the first row in Table~\ref{tab:trainingdata}), and from successful inputs to failure-inducing inputs (the second row in the table). 
%We used both approaches to generate the data.

%\textcolor{blue}{Triggering numerical instability within machine learning models poses a significant challenge due to the complex landscape of high-dimensional input spaces and the non-linear nature of these models. Precisely identifying the threshold at which numerical stability is compromised requires a nuanced approach. Consequently, our mutation strategy incorporates a step-wise increase or decrease, governed by a meticulously chosen "RATE" parameter. This methodical adjustment ensures a deliberate exploration of the input space, enabling the precise identification of inputs that lead to numerical instability. Such a controlled mutation approach is indispensable for generating meaningful training data for Neural Assertions, as it allows for the systematic probing of the model's behavior in response to subtle changes in input, thereby uncovering the delicate balance between stability and instability in numerical computations. }

Here we only used the basic transformations of {\it increasing}, {\it decreasing} and {\it no change} as labels. In the future, we plan to also consider quantized values to provide more fine-grained transformation labels, e.g., {\it increasing 100} is a label, and {\it increasing 200} is another label. This approach potentially leads to more accurate guidance for fuzzers. 

\noindent{\bf (2) Base input generation:} To generate initial base inputs, we sample $n$ values ($n$ is configurable) across different regions of the input space. For example, we can sample 10 inputs less than 0, 10 inputs between (0,100), and 10 inputs larger than 100. If we have the domain knowledge of an unstable function, we can sample regions more efficiently, e.g., directly using some failure-inducing inputs we know.  In our experiments, we set $n$ to 100.

%We generate test inputs by performing mutations on the base inputs. We sampled 100 base inputs at intervals (e.g., (-inf, 0), (0,100), (100, inf)). 

Next, we apply the above mutation techniques to generate new values and test them on the unstable function. As soon as we observe the change of the test outcome for input $x$ and its mutated input $x'$, we record training data points like the ones stated in Table~\ref{tab:trainingdata}. For example, given an initial value 10 (test successes) and its mutated sequence, 20 (test successes), 30 (test successes), 40 (test fails), we collect the training data points (20, increase), (30, increase) and (40, no change) respectively; similarly, given an initial value 100 (test fails), and its mutated sequence, 130 (test fails) and 160 (test successes), we collect the training data point (160, decrease), (100, no change) and (130, no change) respectively. The overall complexity of this process is \( O(n \cdot m) \), where \( n \) is the number of base inputs, and \( m \) is the number of mutations per base input. Our empirical results show that this method can generate failure-inducing inputs for the majority of functions under study. We also added known failure-inducing inputs from the literature \cite{kloberdanz2022deepstability} to increase the chance of triggering failures.

%\noindent{\it Generating Training Data:} 

Our inputs support any size of vectors and matrices. In the experiments, we used randomly generated base inputs as elements of matrices, and we also perturbed the real-world data such as MNIST to obtain inputs. For the real-world dataset, we make sure that the mutated inputs still follow the constraints of image pixels (0,255).

{Sometimes, when we test randomly generated matrices or a perturbed real-world dataset directly against our unstable functions, the tests always fail. In this case, we perform a data pre-processing to scale the data to make sure that some inputs fail, and some are successful. In this way, we can obtain a balanced dataset to train the model. For example, a lot of pixels of the MNIST image have the value 0 (black colored). We added a small $\epsilon$ to ensure when we test an unstable function, it is not always crashing.}

\subsubsection{\textbf{Training a Machine Learning Model to Generate a Soft Assertion}}~\label{model} %\wei{done by wei}
Once we obtain the training data, we train a machine learning model that takes matrix/vectors as inputs and reports the classification specified in formula (1). We explored a variety of machine learning model architectures, including DNN, RNN, LSTM, transformers and K-Nearest Neighbors (KNN) and Random Forest. We found that Random Forest can handle any range of random values without encountering numerical stability itself~\cite{wang2018efficient}. It also achieved highest accuracies for majority soft assertions we generated, without the need of pre-processing of data. See Section~\ref{implementation} for the implementation details of model training.

\section{Fuzzing with Soft Assertions}~\label{sec:fuzzing}
\vspace{-.6cm}
\subsection{The fuzzing Framework} 
\begin{figure}[htbp] % htbp is a placement specifier
  \centering
  \includegraphics[width=0.65\textwidth]{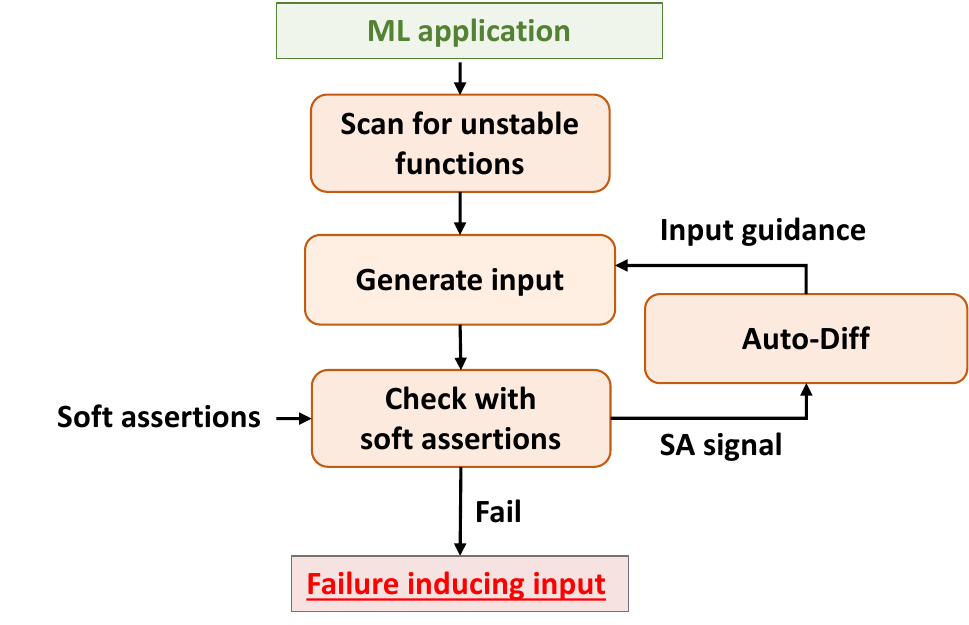}
    \caption{An Overview of \tool}
    \label{fig:framework}
\end{figure}

%\wei{done by Wei}
Figure~\ref{fig:framework} shows an overview of our {\it Soft Assertion Fuzzing} framework. The fuzzer takes an ML application as input. It scans the code and compares our {\it Unstable Function Database} to locate the unstable functions in the ML application. The fuzzer starts with random inputs and during execution of the inputs, it invokes the {\it Soft Assertion Checker} to take the values at the entry of the unstable function and make a prediction using soft assertion.  If the error triggering condition is met, the fuzzer continues the execution to demonstrate the error; otherwise, the {\it Soft Assertion Checker} returns a {\it Soft Assertion (SA) Signal} to teach the fuzzer how to transform the current value to trigger the bug, e.g., increasing or decreasing the values. This proposed transformation is propagated to the input via the {\it Auto-diff} component to guide the mutation to generate the input for next iteration.

%The input generation component will also leverage the  the history of the input and its mutations and will generate the input based on the guidance and the history.

% \begin{figure}[htbp] % htbp is a placement specifier
%   \centering
%   \includegraphics[width=0.7\textwidth]{Images/framework.png}
%     \caption{An Overview of \tool}
%     \label{fig:framework}
% \end{figure}

\vspace{-12pt}
\subsection{The Fuzzing Algorithm}
%\wei{done by wei}
In Algorithm \ref{alg:neural_assert_fuzzer}, we present our fuzzing technique that uses soft assertions.  The input of our tool is (1) a machine learning application, $p$, (2) the unstable function database $D$ (Section~\ref{database}), and (3) their soft assertions, $A$, we trained for the unstable functions (Sections~\ref{oracle}----\ref{model}).  The output is the failure-inducing inputs, $I$, that can trigger numerical bugs for the unstable functions in $p$.%, as well as the time used to generate the inputs, $t$.

%In this section, we introduce our Neural Assertion Fuzzing technique, which integrates Neural Assertions (NA) with a specialized fuzzing methodology to enhance the detection and resolution of numerical bugs in machine learning (ML) applications. Our approach, as outlined in Algorithm \ref{alg:neural_assert_fuzzer}, is designed to leverage the precision of Neural Assertions alongside the exploratory nature of fuzzing, allowing for more effective navigation through the complex input spaces of ML code. This fusion of methodologies enables us not only to pinpoint numerical bugs with greater accuracy but also to do so in a manner that is both systematic and efficient.

\begin{algorithm}
\caption{Soft Assertion Fuzzing Algorithm}\label{alg:neural_assert_fuzzer}

\textbf{Input:} Program $p$, Unstable function database $D$, Soft Assertions $A$

\textbf{Output:} Failure-inducing inputs $I$%, Time used $t$
 
\begin{algorithmic}[1]

\State $I \leftarrow \emptyset$ % Initialize set of failure-inducing inputs
\State $F \leftarrow$ ScanForUnstableFunctions$(p, D)$
%\State Initialize timer with threshold $T_{max}$

\For{$f \in F$}
    \State ${x} \leftarrow$ 
    GenerateInitialInput$()$
    \State ${H} \leftarrow \emptyset$
    \While{ $t$ $<$ $T_{timeout}$}
        \State ${v} \leftarrow$ Execute$({x}, p, f, \text{entry})$
        \State $signal \leftarrow$ SoftAssertion$(A, {v}, f)$
        \If{$signal == \text{NO CHANGE}$}
            \If{Execute$({x}, p, f, \text{exit}) == \text{SAFE}$}
                ${x} \leftarrow$ GenerateInitialInput$()$
            \Else           
                \State {add ${x}$ to $I$; print $t$;  \textbf{break}}
            \EndIf
        \Else
            \State $\Delta x \leftarrow$ AutoDiff$({x}, f, signal)$         
            \State ${x'} \leftarrow$ ConstraintSolving$({x}, \Delta x, H)$
            \State add $\left<x, \Delta x\right>$ to $H$; $x\leftarrow x'$        
        \EndIf
    \EndWhile
\EndFor

\State \Return $I$
\end{algorithmic}
\end{algorithm}

At line~2, we first scan the code to identify unstable functions in program $p$. For each identified unstable function $f$, we search for inputs that can lead to its failure starting at line~3. At line~4, we first randomly generate an initial input for ${x}$, and at line~6, the algorithm iterates throughout a loop, where we mutate the input in successive iterations to search for a failure-inducing input. $T_{timeout}$ is a configurable timer we used to control fuzzing time. At line~5, we initialize $H$, which will store the history of the inputs tried during fuzzing. In particular, we store the inputs at each iteration together with the update instructions we propagated from soft assertion signals as well as the test outcomes. This history provides us constraints to search for inputs of next iterations. For example, from the history (10, increase, success) and (100, decrease, success), we can obtain the constraints to generate failure inducing inputs $10<x<100$, and then we can use constraint solver to generate the input.

%Following this initialization, the algorithm iterates within a loop limited by a predetermined time threshold $T_{max}$, initiated at Line 3. Throughout this loop, the focus is on mutating the input in successive iterations to discover a failure-inducing input. 
%. This evaluation $\mathbf{v}$ is then processed through our Neural Assertion $A$, resulting in a signal, shown in Line 9. 

At line~7, we run the current input $x$ on $p$ till it reaches the entry of $f$, where we record the value $v$. At line~8, we take this value and query the soft assertion of $f$. If the returned soft assertion signal suggests a failure with the label {\it no change} (see line~9), we perform a validation by continuing executing $x$ on $p$ till the end of $f$. At line~10,  we compare the output at the exit of $f$ with the oracle. If the oracle reports that the current input $x$ is safe, indicating a misprediction by the soft assertion, the process is reset with a new initialized input. However, if the oracle confirms that the execution fails with $x$, we add $x$ to the failure-inducing input set $I$, and print the time used, $t$, at line~12. This concludes the search for the current function $f$, and we then move on to the next unstable function in $F$.%, at line~3.

If at line~9, the soft assertion does not return {\tt NO CHANGE}, we will need to mutate the current input $x$ towards the areas more likely to uncover failure-inducing inputs. The soft assertion signals are designed for providing such feedback. At line~15, we propagated the soft assertion signal generated at $f$ back to the entry of the program using an auto differentiation component (details in Section~\ref{idea}).
The adjustment $d$ computed from line~15 will be used to update the current input $x$ to generate a new input $x'$. In the process of updating the current input $x$, shown at line~16, we incorporate the history of previous inputs, denoted by $H$, which is initially set up at line 5 (details in Section~\ref{idea}). {Subsequently, at line 17, both $x$ and $d$ are added to the input History $H$, and the new input $x'$ is set to $x$, and will continue in next iteration and be executed at line~7.} %This recorded history is then leveraged in subsequent iterations to guide the mutation more effectively. 

%This utilization of input history allows for more informed decisions in the mutation process
%using the signal and auto differentiation, with the specific calculation detailed in Equation \ref{eq:1}. Following this, the current input $x$ is updated accordingly, with the mutation details provided in Lines 18-19. In the process of updating the current input $x$, we incorporate the history of previous inputs, denoted by $H$, which is initially set up in Line 6. This utilization of input history for adaptive mutation allows for more informed decisions in the mutation process. Subsequently, in Line 20, both $x$ and $d$ are added to the Input History $H$. This recorded history is then leveraged in subsequent iterations to guide the mutation process more effectively. 

\subsection{Using Soft Assertion Signals to Guide Input Generation}~\label{idea}
%\wei{done by wei}
Soft assertion signals (generated at line~8 in Algorithm~\ref{alg:neural_assert_fuzzer}) instruct, at the assert location, how to adjust the current values to trigger the instability in the unstable function. We formulate the details of our approach in math as follows. Given an input $x$ and a section of code, annotated as $g$, before reaching the unstable function $f$, the value $y$ at the entry of $f$ is $g(x)$. We can use {\it auto-differentiation}~\cite{baydin2018automatic} to generate the output of $g'(x)$, where $g'$ is the derivative function of $g$.
$g'(x) = \Delta y/\Delta x$ captures the rate of the change of $y$ around a particular value of $x$. When the soft assertion outputs a signal on how to transform $y$ to trigger instability, i.e., $\Delta y$, we can then compute the corresponding update of input $\Delta x$  using $\Delta y/g'(x)$. 

Considering training a machine learning model to predict the concrete value of $\Delta y$ can be hard, in this paper, we explore the feasibility and effectiveness of using the directions of value changes, such as "increase" and "decrease"; that is, in the {\it AutoDiff} function in Algorithm~\ref{alg:neural_assert_fuzzer} at line~15, we implemented the following formula:

\begin{equation} \label{eq:1}
\Delta \mathbf{x} = \frac{{SA(X)} \times \text{Rate}}
{{\nabla_{\mathbf{x}} g(\mathbf{x})}}
\end{equation}

Here, $\Delta \mathbf{x}$ represents the adjustment to program input $\mathbf{x}$, $SA(X)$ is the direction suggested by the soft assertion defined in formula (1),
%---if the soft assertion signal is "increase", we set $SA(X)$
% $\text{Signal}_{\text{NA}}$ to positive, and otherwise, negative
$\nabla_{\mathbf{x}} g(\mathbf{x})$ is the gradient from automatic differentiation (AD), and \textit{Rate} adjusts the mutation's intensity, which is configurable in our tool.  

%The input update direction is determined by the interaction between the soft assertion signal $SA(X)$ and the gradient direction determined by formula (\ref{eq:1}) as shown above. When SA(X) is positive ("increase") and the gradient direction is positive, the input x should change in the same direction to move the value y in the unstable function toward potential instability. However, when SA(X) is positive but the gradient direction is negative, the input x should be adjusted in the opposite direction to still achieve an increase in y.In contrast, when SA(X) is negative ("decrease") and the gradient direction is positive, the input x should be decreased, whereas if the gradient is negative, the input x must increase. Additionally, when SA(X) returns "no change", this indicates that the current values are in a potentially unstable state and no further adjustments are needed.

% {Table \ref{tab:input_update} further clarifies how a soft assertion signal interplays with the gradient direction reported by AD in determining the direction of input value adjustments. 
% For example, the first row indicates that
% When $SA(X)$ is positive ("increase") and the gradient direction reports positive, the input $x$ should change the same direction as the value $y$ at the unstable function, which is also "increase". \textcolor{red}{Similarly, when $SA(X)$ is negative and gradient direction is positive, the input $x$ should change the direction to "decrease" as the value $y$ at the unstable function}
% The final mutation values are determined by formula (\ref{eq:1}) as shown above.

\noindent{\bf Leveraging Input History for Mutation:} In Algorithm~\ref{alg:neural_assert_fuzzer} at line~16, we used the history of generated inputs to help determine the values of a new input. Specifically, we construct a set of constraints based on the pairs of $\left<x,d\right>$ stored in $H$ at line~18. For example, for $\left<30, decrease\right>$, we add $x<30$;  for $\left<-1, increase\right>$, we add $x>-1$. We combine these constraints with formula~(\ref{eq:1}). to generate a new input.% $x$ generated from Eq.1 to the constraints when generating a new input.% \textcolor{blue}{The \textbf{Up-Down-Up Strategy} is a paradigm we've named and implemented in our algorithm, reflecting a systematic approach to refining input mutations. This strategy, embedded within our method, leverages historical input data to guide the mutation process. If an input adjustment does not yield the desired outcome, we reverse the change direction and adjust again, aiming for a more nuanced approach to trigger potential issues. This iterative feedback loop, inspired by natural selection processes, allows our algorithm to 'learn' from previous mutations, enhancing its efficiency and precision in identifying errors within machine learning applications.}

\section{Evaluation}~\label{sec:results}
To evaluate our {\it Soft Assertion Fuzzer}, we studied the following research questions:
\begin{itemize}
\item {\bf RQ1:} Can we effectively generate soft assertions?
\item {\bf RQ2:} Can our \tool effectively detect numerical bugs?
\item {\bf RQ3:} {In what aspects, does \tool perform better than the state-of-the-art techniques?}
\item {\bf RQ4:} How do different configurations affect the effectiveness of {\it soft assertion fuzzer} (ablation studies)?
\end{itemize}

\subsection{{Experimental Setup}}
%\wei{done by wei}
\subsubsection{\bf Implementation}~\label{implementation}
We implemented our \tool in Python 3.10.12, the most stable and recent version.
It consists of a lightweight static analysis tool that scans the code and compares it with the unstable function database to locate unstable functions in the code. It invokes soft assertions during fuzzing in these unstable functions. We currently can handle programs that use PyTorch or Tensorflow libraries. We used the {\it torch} {\it autograd} package, which can handle ML applications that use {\it torch} libraries, and {\it Tensorflow}'s autodiff methods, called {\it gradients}, which can handle Tensorflow libraries. We conducted our experiments on Python benchmarks, but we believe that with small modifications, \tool  can also work for Java or R programs.

Out of 61 unstable functions in the database, we implemented soft assertions for 25 representative functions. We included {all the unstable functions GRIST~\cite{yan2021exposing}handled, including {\tt log}, {\tt sqrt} and {\tt exp}}, so that we can compare the GRIST benchmark. We ensure that the 25 unstable functions cover a variety of types, from mathematical operations (e.g,  {\tt reciprocal} and {\tt division}), activation functions (e.g, {\tt softmax}, {\tt sigmoid} and {\tt relu}), to hyperbolic functions (e.g, {\tt tanh} and {\tt sinh}), so that our tool can benefit a variety of ML applications. To demonstrate the advantage of soft assertions, we especially sampled the functions, e.g., {\tt matmul} and {\tt mean}, where we do not have knowledge on their safety input conditions, and it is very hard to add a traditional symbolic assertion to protect those functions. To evaluate whether the ML models are capable of encoding error conditions of complex math functions, we considered a variety of math functions frequently used in the neural networks, including the inverse trigonometric functions (e.g, {\tt acos}), exponential operations (e.g,  {\tt power} and {\tt exp}), {\tt cosine similarity} and {\tt cross entropy} as well as {\tt linear} and {\tt conv2d}.

%\wei{till here}
%Unlike other models Random Forest does not need explicit loss function or optimizer. In addition, RandomForest model is less sensitive in hyper-parameter tuning, we found 100 decision work best in our dataset.
%for the most of the unstable functions
To generate soft assertions, we implemented a set of neural network architectures and ML
models such as DNN, RNN, LSTM, and transformers using PyTorch, as well as  KNN and Random Forest using Scikit-learn. We did pilot studies with these models on a subset of unstable functions under study, and found that Random Forest performed the best. Specifically, we used RandomForestClassifier from Scikit-learn with 100 decision trees, random selection of data subsets is 42. For each unstable function, we trained a model, using automatically generated datasets of the size 40 k (30\% of these datasets are used for testing). We explored several settings of input data, including the matrices of $3\times3$, the matrices of $14\times14$ (half of the size of an MNIST image) as well as MNIST.  We also explored, in our pilot study, to use the CIFAR and Fashion-MNIST datasets for unstable functions, but we found that the model accuracies are not as good compared to smaller matrices.
We trained each model for each dataset three times and logged the average F1 score and training time.
Note that model training is a one-time process, and once done, it can be repeatedly used in different ML applications. %Apart from the MNIST dataset , we have tested CIFAR image dataset for unstable functions Softmax and acos in RandomForest model with F1 score of  0.6759 and 0.6238 respectively. In addition we did the same experiment with Fashion-MNIST image dataset for Softmax and acos with F1 score of 0.6558 and 0.6317 respectively.
%We used the Adam optimizer and set the number of training epochs to 100 for all models. Additionally, we set the learning rate to 0.002. For the RNN and LSTM models we added drop out layers with the dropout rate of 0.5. For the Random Forest model, we optimized performance using 100 trees, and for the KNN model, we selected the 5 nearest neighbors for predictions. We trained the models on automatically generated datasets of the size 40 k (see Section~\ref{train} for the approach) and used 30\% of these datasets for t{esting and checking the models' performance. \}textcolor{blue}{[SB] In order to generate neural assertions, we implemented the Random Forest models using sklearn and PyTorch.
\subsubsection{\bf Baselines} We used five SOTA of Python fuzzers as baselines, including {\it GRIST}~\cite{yan2021exposing} and {\it RANUM} ~\cite{li2023reliability} targeting numerical bugs in ML applications, {\it PyFuzz}~\cite{pyfuzz2022}, a Python fuzzer released in 2022, {\it Atheris}\cite{atheris2023}, a well-maintained fuzzer developed by Google for coverage-guided fuzzing for Python, and {\it Hypothesis}~\cite{hypothesisDocs}, a widely used framework for property-based testing.
%\textcolor{orange}{Additionally, we included {\it RANUM} ~\cite{li2023reliability},the most recent approach, which builds on the GRIST codebase to enhance numerical bug detection in python-based machine learning applications.}

We used the default settings and hyper-parameters to run the baselines. We ran these fuzzers 5 times for each program and get an average time as the results. We set the timer for 30 minutes for each program for all the fuzzers including ours.

%To evaluate our techniques, we constructed a benchmark consisting of {30} ML programs, among which, {19} are from the GRIST paper, where we knew existing numerical bugs (it is useful to evaluate the false negatives of our tool), and 11 are from the popular GitHub projects where we don't know the bugs (it can help evaluate the usefulness of our tool for real-world code). 

%{To the best of our knowledge, the GRIST benchmark is the only existing benchmark for numerical bugs in ML applications. It has been used by other recent tools, e.g., RANUM~\cite{li2023reliability}.

\subsubsection{{\bf Constructing a benchmark of ML applications}}~\label{benchmark} {To evaluate  our techniques, we constructed a benchmark consisting of {94} ML programs, among which, {79} programs are from the GRIST benchmark~\cite{yan2021exposing}, where we know existing numerical bugs (useful to evaluate the false negatives of our tool). GRIST has also been used by evaluating other recent tools, e.g., RANUM~\cite{li2023reliability}. 

Additionally, we selected {15} programs from various GitHub projects where we don't know the bugs, which helped evaluate the usefulness of our tool for real-world code.} To find important and recent projects in GitHub, we used GitHub API to search for Python code. Our query targets repositories using the NumPy, TensorFlow or PyTorch framework, focusing on the implementation that used unstable functions by providing these keywords such as {\tt sqrt}, {\tt exp} and {\tt Softmax}. From the search results, we selected 30 repositories that have at least 10 stars to make sure we focus on potentially impactful projects. We manually reviewed these repositories and kept the ones that have been updated within the last three years. We also excluded the ones that potentially use incompatible Python versions and obsolete frameworks. Sometimes, the given keyword is mentioned in the comment section but not implemented in the code; we exclude those cases as well.  After filtering, we obtain a total of 15 programs as the subjects for our evaluation.

\subsubsection{\bf Running Experiments}
All of our soft assertions are trained on a single NVIDIA (A100-SXM4-80GB) GPU, supported by an x86\_64 architecture CPU with 251.35 GB of RAM. We conducted our fuzzing experiments on the Intel(R) Core(TM) i5-8365U CPU @ 1.60GHz with 16GB RAM, Ubuntu 22.04.6 LTS. We used the Anaconda environments to switch between versions of PyTorch and TensorFlow for the benchmark programs that use different library versions.
\begin{table}
    \centering
    \caption{Generating soft assertions: F1 scores and training time in minutes}
    \footnotesize
\begin{tabular}{clcccccc}
\toprule
\textbf{No.} & \textbf{Unstable Func} & \textbf{$3\times3$ matrix} & \textbf{Train time} & \textbf{$14\times14$ matrix} & \textbf{Train time } & \textbf{MNIST} & \textbf{Train time} \\
\midrule
1  & Softmax          & \textbf{0.8145} & 0.286    & 0.8106  &  0.998  & 0.7107     & 0.796 \\
2  & log              & \textbf{0.6574} &  0.282   & 0.6172   & 1.149  & 0.5018     & 0.320 \\
3  & sigmoid          & \textbf{0.6210} &  0.360  & 0.6160   &  1.188 & 0.5039     & 0.948 \\
4  & exp              & 0.6663  & 0.115  & \textbf{0.8101}  &  1.001  & 0.7038     & 0.804 \\
5  & logSftmx         & 0.5518  & 0.481  & 0.6289 &    1.274  & \textbf{0.7136}     & 0.776 \\
6  & sqrt             & 0.6134  & 0.319  & 0.6343 &    1.645 & \textbf{0.8011}     & 3.567 \\
7  & tanh             & 0.5411  & 0.349  & 0.5334 &    1.688 & \textbf{0.7241}     & 0.756 \\
8  & ReLU             & 0.6136  & 0.352  & \textbf{0.6171}  &  1.709  & 0.5424     & 1.105 \\
9  & ELU              & \textbf{0.6469}  & 0.348  & 0.6421  &  1.900  & 0.5441     & 0.761 \\
10 & SoftPlus         & 0.6228  & 0.370  & \textbf{0.6267}  &  1.141  & 0.6034     & 2.482 \\
11 & rSqrt            & \textbf{0.6204}  & 0.392  & 0.5977  &  4.869  & 0.5295     & 1.099 \\
12 & Div              & 0.6215  & 0.359  & \textbf{0.6255}  &  1.803  & 0.5426     & 0.470 \\
13 & linear           & \textbf{0.8141}  & 0.241  & 0.7870  &  1.422  & 0.7141     & 1.378 \\
14 & matmul           & 0.6657  & 0.298  & 0.6562 &    1.759 & \textbf{0.7134}     & 1.379 \\
15 & mean             & 0.5272  & 0.255  & \textbf{0.6661}  &  1.353  & 0.5524     & 1.262 \\
16 & reciprocal         & 0.6376  & 0.380  & \textbf{0.6427}   &  2.050 & 0.5630     & 0.375 \\
17 & CosSim           & 0.6503  & 0.342  & \textbf{0.6632}   & 1.366  & 0.5062     & 0.394 \\
18 & acos             & 0.6681  & 0.241  & 0.6673   & 1.278  & \textbf{0.6810}     & 0.416 \\
19 & cosh             & 0.8092  & 0.277  & \textbf{0.8133}  & 0.980   & 0.7644     & 0.675 \\
20 & sinh             & 0.8119  & 0.279  & \textbf{0.8134}  &  0.993  & 0.5094     & 1.248 \\
21 & square           & \textbf{0.6702}  & 0.243  & 0.6631   & 1.397  & 0.5028     & 1.215 \\
22 & pow              & \textbf{0.6667}  & 0.240  & 0.6590   &  1.318 & 0.5549     & 2.645 \\
23 & sum              & 0.6601  & 0.254  & \textbf{0.6689}  &  1.290  & 0.5601     & 1.761 \\
24 & CE               & 0.6380  & 0.370  & \textbf{0.6447}  &  1.981  & 0.5472     & 1.329 \\
25 & Conv2d           & 0.6347  & 0.372  & \textbf{0.6415}   & 1.945  & 0.5396     & 1.340 \\
\midrule
   & \textbf{Average} & \textbf{0.6577}  & 0.312  & \textbf{0.6698}  & 1.580   & \textbf{0.6051}     &   1.172   \\
\bottomrule
\end{tabular}

    \label{tab:f1_score}
     \vspace{-.39cm}
\end{table}

\subsection{RQ1: Generating Soft Assertions by Testing Unstable Functions}
%\wei{done by wei}

%we show the results where we used the $3\times3$ and $14\times14$ matrices as input and followed Section~\ref{train} to generate the data. Under \textit{MNIST}, we used perturbed MNIST matrices as input. 
In Table~\ref{tab:f1_score}, we report our results on soft assertion generation for 25 unstable functions. We used three sizes of input, including {\it $3\times3$ matrix} and {\it $14\times14$} matrix as well as the perturbed MNIST ({\it $28\times28$} matrix), following Section~\ref{train}. Across 25 unstable functions listed in the second column, we achieved the average F1 scores of 0.6577, 0.6698 and 0.6051 respectively for the three settings. Some soft assertions are accurate and achieved 0.8145 F1 score (Column {\it $3\times3$ matrix} Row {\it Softmax}) and 0.8101 F1 score (Column {\it $14\times14$ matrix} Row {\it exp}). We did not observe the advantages of one setting over the others. The accuracy seems to be dependent on individual unstable functions. Although the F1 scores of these soft assertions are not always very high, we found that they already are very useful for helping fuzzers trigger numerical bugs (see Sections~\ref{fuzzing} and ~\ref{sec:baseline}). We also reported the training cost for each unstable function. The last row in Table~\ref{tab:f1_score} shows that the models were trained fast and can finish 0.312--1.580 minutes on average for the three types of input data.

\subsection{RQ2: Detecting numerical bugs in ML 
applications}~\label{fuzzing}

\begin{table}[ht]
    \centering
    \caption{Detect numerical bugs for real-world programs: (1) compare with baselines, (2) ablation studies}
    \resizebox{\textwidth}{!}{
    \label{tab:baseline_vs_soft_assert}
    \begin{tabular}{|c|p{3.2cm}<{\raggedright}||c|c|c|c|c||c||c|c|c|c|c|c|}
    \hline
    \begin{tabular}[c]{@{}c@{}} \textbf{Github} \\ \textbf{Repo} \end{tabular} & \textbf{Failure Symptoms} & \multicolumn{5}{c||}{\textbf{Comparing with Baselines (Time in seconds)}} & \begin{tabular}[c]{@{}c@{}} \textbf{SA} \\ \textbf{Fuzzer}\end{tabular}
     & \multicolumn{6}{c|}{\textbf{Ablation Study}} \\ \cline{3-7} \cline{9-14}
    & & \textbf{PyFuzz} & \textbf{Hypo} & \textbf{Atheris} & \textbf{GRIST} & \textbf{RANUM} & & \textbf{No $H$} & \textbf{MNIST} & \textbf{DNN} & \multicolumn{3}{c|}{\textbf{3} $\times$ \textbf{3}} \\ \cline{12-14}
    & & & & & & & & & & & \textbf{40K} & \textbf{15K} & \textbf{5K} \\ \hline\hline

    ~\cite{ID_5} & NaN in {\tt rqst}: negative pixel values after normalization
    & $\times$ & $\times$ & $\times$ & $\times$ & $\times$ & \textbf{0.0243} & $\times$ & 0.3320 & 0.4200 & 0.0645 & 0.1289 & 0.1935 \\ \hline
    ~\cite{ID_2} & $\times$ & $\times$ & $\times$ & $\times$ & $\times$ & $\times$ & $\times$ & $\times$ & $\times$ & $\times$ & $\times$ & $\times$ & $\times$ \\ \hline
    ~\cite{ID_3} & $\times$ & $\times$ & $\times$ & $\times$ & $\times$ & $\times$ & $\times$ & $\times$ & $\times$ & $\times$ & $\times$ & $\times$ & $\times$ \\ \hline
    ~\cite{ID_4} & INF in {\tt exp}& $\times$ & 0.0438 & $\times$ & 140.8300 & 1.4400 & 0.0660 & 0.0070 & \textbf{0.0070} & 0.0123 & 0.0070 & 0.0133 & 0.0210 \\ \hline
    \cite{ID_1} & $\times$ & $\times$ & $\times$ & $\times$ & $\times$ & $\times$ & $\times$ & $\times$ & $\times$ & $\times$ & $\times$ & $\times$ & $\times$ \\ \hline
    ~\cite{ID_6} & NaN in {\tt power} caused by overflow& $\times$ & $\times$ & $\times$ & $\times$ & $\times$ & 0.0150 & 0.0090 & 0.0084 & 0.0139 & \textbf{0.0080} & 0.0159 & 0.0253 \\ \hline
    ~\cite{ID_7} & -INF in {\tt log} for very small probabilities (close to zero)& $\times$ & $\times$ & $\times$ & 18.6500 & 0.6180 & 0.0070 & \textbf{0.0030} & 0.0170 & 0.0192 & 0.0170 & 0.0323 & 0.0510 \\ \hline
    ~\cite{ID_8} & NaN in {\tt matmul} caused by overflow& $\times$ & $\times$ & $\times$ & $\times$ & $\times$ & \textbf{0.1840} & 1.1300 & 0.9440 & 1.4500 & 0.9400 & 1.7920 & 1.8320 \\ \hline
    ~\cite{ID_9} & NaN in {\tt L2 norm} caused by overflow& $\times$ & $\times$ & $\times$ & $\times$ & $\times$ & \textbf{0.0100} & 0.9000 & 1.7950 & 2.0500 & 1.7000 & 3.4105 & 3.4565 \\ \hline
    ~\cite{ID_10} & INF in {\tt matmul}  & $\times$ & $\times$ & $\times$ & $\times$ & $\times$ & \textbf{0.2480} & 1.3000 & 1.2700 & 1.1400 & 1.2700 & 2.4110 & 2.8580 \\ \hline
    \multirow{2}{*}{~\cite{ID_11}} & INF in {\tt exp} & $\times$ & $\times$ & $\times$ & $\times$ & $\times$ & \textbf{17.0500} & 27.7600 & 32.0400 & 42.1000 & 32.0500 & 60.8950 & 87.8950 \\ \cline{2-14}
    & Inconsistency in PyTorch for {\tt cosine similarity} & 2.9010 & $\times$ & 2.9480 & $\times$ & $\times$ & \textbf{0.0010} & 0.0300 & 0.0020 & 0.0033 & 0.0020 & 0.0038 & 0.0061 \\ \hline
    ~\cite{ID_12} & Invalid input exception in {\tt Root Mean Squared} & 0.0126 & 7.1700 & 0.0685 & $\times$ & $\times$ & \textbf{0.0030} & 0.0030 & 0.0030 & 0.0048 & 0.0030 & 0.0057 & 0.0092 \\ \hline
    ~\cite{ID_13} & NaN in {\tt softmax} & $\times$ & $\times$ & $\times$ & 4.2940 & 0.7060 & \textbf{0.6920} & 0.7040 & 0.8830 & 1.1000 & 0.7160 & 1.6777 & 2.6550 \\ \hline
    ~\cite{ID_14} & Incorrect value in {\tt linear solver} & $\times$ & $\times$ & $\times$ & $\times$ & $\times$ & \textbf{1.0200} & $\times$ & 1.4450 & 4.0200 & 2.7760 & 2.7455 & $\times$ \\ \hline
    ~\cite{ID_15} & Wrong prediction due to vanishing gradient & 0.7069 & 32.4000 & 0.4121 & $\times$ & $\times$ & \textbf{0.6260} & 0.9340 & 0.9630 & 1.5020 & 0.9220 & 1.8285 & 2.8920 \\ \hline\hline

    \multicolumn{2}{|c||}{{\bf Total Bugs}} & 3 & 3 & 3 & 3 & 3 & \textbf{13} & 11 & 13 & 13 & 13 & 13 & 12 \\ \hline
    \multicolumn{2}{|c||}{{\bf Average Time}} & 1.2068 & 13.2043 & 1.1432 & 54.3580 & 0.9213 & \textbf{1.9182} & 3.0555 & 4.0453 & 4.1412 & 4.6637 & 5.7662 & 8.4913 \\ \hline

    \end{tabular}
  }  
\end{table}

\begin{table}
\caption{Outperform the State-of-the-Art Techniques}\label{tab:baseline} 
\centering
\resizebox{0.9\textwidth}{!}{%
\begin{tabular}{|l||c|c||c|c|c|c|c||c|c|c|}
\hline
Fuzzer & GRIST (79) & GRIST Average Time (sec) & Real-World (15) & NaN/INF & Other Failures\\\hline\hline
SA Fuzzer & 79 & 0.646 & 13 & 88 & 4\\\hline
RANUM~\cite{li2023reliability} & 79 & 2.209 & 3 & 82 & 0\\\hline
GRIST~\cite{yan2021exposing} & 78 & 44.267 & 3 & 81 & 0\\\hline
Atheris~\cite{atheris2023} & 25 & 0.283 & 3 & 27 & 1\\\hline
PyFuzz~\cite{pyfuzz2022} & 24 & 5.498 & 3 & 26 & 1\\\hline
Hypothesis~\cite{hypothesisDocs} & 23 & 5.945 & 3 & 25 & 1\\\hline
\end{tabular}
} 
\end{table}

\vspace{-.3cm}
\tool found 13 bugs from 15 real-world programs, shown in  Table~\ref{tab:baseline_vs_soft_assert}, and found all the GRIST bugs (79 total, see the first row in Table~\ref{tab:baseline}). We listed the links of the GitHub repositories, and reported the failure symptoms in the first and second columns of Table~\ref{tab:baseline_vs_soft_assert}. Column {\it SA Fuzzer} reports our results using the soft assertion trained with the $14\times14$ dataset. This dataset reported the best average accuracy in Table~\ref{tab:f1_score}.  Each tabular reports the time used to trigger the bug. If the bug is not found, we mark an $\times$. Since the GitHub bug is unknown, an $\times$ could mean that there does not exist a numerical bug in the application. Our results show that finding these bugs are very fast in seconds ({1.92 seconds} on average) using our SA Fuzzer. Under {\it Failure Symptoms}, we show that 
we can detect NaN/INF errors, as well as incorrect results.  Under {\it Comparing with Baselines}, we listed 5 baseline results for the real-world bugs, which we will discuss in more details in Section~\ref{sec:baseline}.

\begin{wrapfigure}{l}{0.56\textwidth} 
  \centering
  \includegraphics[width=0.56\textwidth]{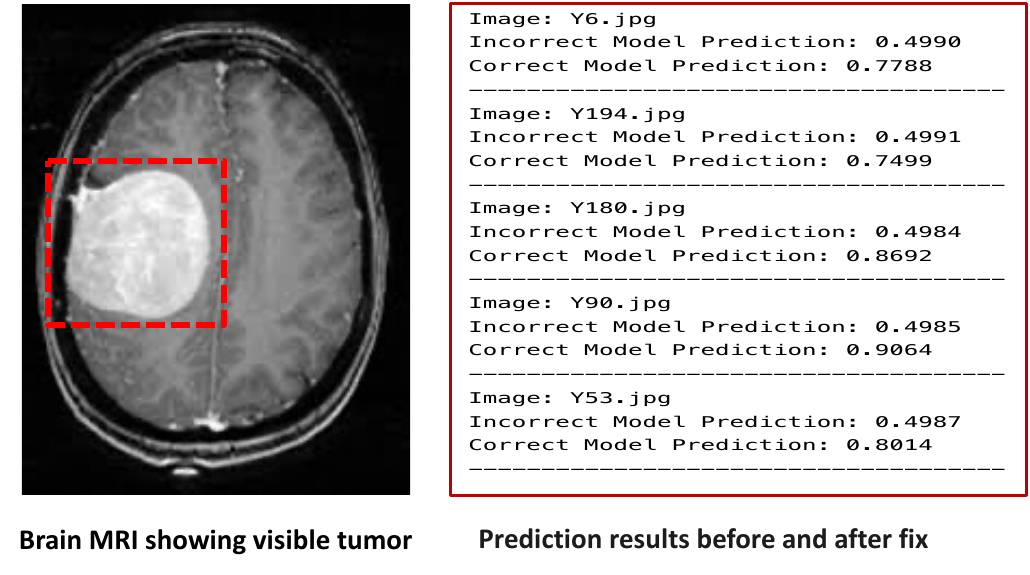}
  \caption{Numerical instability leads to model mispredictions}
  \label{fig:casestudy}
\vspace{-.4cm}
\end{wrapfigure}

\noindent{\bf Case Study: Finding an unknown bug in a tumor detection model.}
 One of the numerical instability we found is in TumorScope~\cite{ID_15}, an ML model trained for tumor detection in brain MRI images using Kaggle dataset~\cite{MRI_Images} (see Table ~\ref{tab:baseline_vs_soft_assert}, the case with vanishing gradient~\cite{ID_15}). Due to this bug, some images with tumor (shown the left in Figure~\ref{fig:casestudy}) are incorrectly predicted with no tumor. 
 %TumorScope, consists of 6 convolutional layers, 6 max pooling layer, 4 dense layers, and 3 dropout layers with ReLU and Sigmoid activation. 

\begin{figure}[!ht]
\centering
\includegraphics[width=\linewidth]{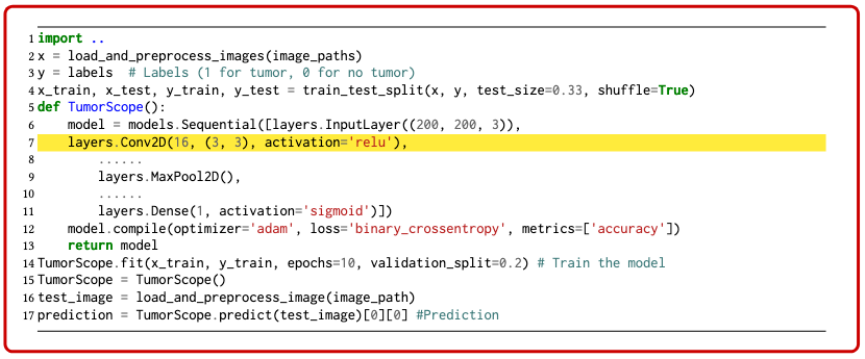}
\caption{TumorScope Model}  \label{fig:case_study_code}
\end{figure}

We observed that during inference, these input images triggered numerical instability in the {\tt Conv2D} layer, shown at line 7 in Figure~\ref{fig:case_study_code}, and produced very small or negative values. The {\tt ReLU} activation function, applied after the {\tt Conv2D} layer, sets all negative values to zero. This results in many neurons outputting zero. As these zeros passed to the subsequent {\tt MaxPool2D} layer (line 9), the layer again propagates mostly zeros. MaxPooling selects the maximum value from each window size of the feature map, so if most inputs to the MaxPooling operation are zero, the output will also be predominantly zeros. Therefore, at the dense layer (line 11), the features extracted from the image are mostly irrelevant, causing the model to produce an incorrect prediction in the final output layer (line 17). 
 We found that the numerical instability was triggered in {\tt Conv2D} because it includes an unstable weight initialization function {\tt GlorotUniform}; as a result, the weights are not properly initialized during training and lead to small/negative values in ReLU activations, causing classification errors.
 When we use a stable weight initialization algorithm, namely {\tt he\_normal}\cite{he2015delving}, and add it at line~7 in Figure~\ref{fig:case_study_code}, the numerical instability is then  fixed.  The  model correctly classified the same tumor image on the left in Figure~\ref{fig:casestudy} with a confidence score of 0.8874 and also correctly predicted other images, shown on the right in Figure ~\ref{fig:casestudy}.

%\vspace{-0.2cm}

\subsection{RQ3: Outperform the Baselines}~\label{sec:baseline}
In Table~\ref{tab:baseline}, we summarized our comparison with the baselines. Because we enabled automatic learning the error conditions in soft assertions, we supported a more variety of unstable functions; that allows us to more effectively find bugs in real-world code, shown in Table~\ref{tab:baseline_vs_soft_assert} under {\it Failure Symptoms.} Among the 15 real-world projects, \tool detected 13 bugs while other baselines all detected 3, shown in Table~\ref{tab:baseline} under {\it Real-World}. We developed a comprehensive set of oracles to capture the failure symptoms beyond {\it NaN/INF}. See the total failures and types generated under {\it NaN/INF} and {\it Other Failures} in Table~\ref{tab:baseline} for both GRIST and real-world programs. %Sometimes in correct values can trigger exceptions in code, see Row 12 in Table~\ref{tab:baseline_vs_neuro_assert}, which are the cases under {\it other failures} for our baselines 

Our results also show that soft assertion guidance effectively helped trigger the bug.  On the GRIST benchmark containing 79 known numerical bugs, \tool matched RANUM's~\cite{li2023reliability} performance by detecting all 79 bugs (compared to GRIST's 78). Although GRIST and RANUM both can handle {\tt exp}, only our tool found the INF bug in {\tt exp} function reported in ~\cite{ID_11} in Table~\ref{tab:baseline_vs_soft_assert}. We believe that the soft assertion guidance also helps quickly trigger the bugs. SA Fuzzer reported 0.646 seconds under {\it GRIST average time}.

\subsection{RQ4: Ablation Study}
%\textit{Experimental Datasets, Settings and Hardware \& Runtime Environments } are same as RQ2.
%In Table~\ref{tab:baseline_vs_neuro_assert} under {\it Ablation Study}, 

We studied whether the use of history inputs (see Section~\ref{idea}) helped performance. We ran \tool by turning off the history information and reported the results under \textit{No H} in Table~\ref{tab:baseline_vs_soft_assert} under {\it Ablation Study}. Our results show that without using history information, we only found 11 bugs, instead of 13, in real-world programs, and it took longer to find the bugs.

We also reported fuzzing results that used soft assertions trained with different sizes of input matrices. See Columns {\it 3x3 } and {\it MNIST} in Table~\ref{tab:baseline_vs_soft_assert}.
Interestingly, SA Fuzzer found the same 13 bugs in the real-world programs in these configurations. We observed that a soft assertion trained with $3\times3$ input matrices can handle ML applications that take the real-world datasets, such as MNIST, as input. We see that  {\it $14\times14$} configuration (under Column {\it SA Fuzzer}) reported the fastest time for most of the bugs. We believe that the efficiency is related to the model accuracy. When we have low mispredictions in soft assertions, the search of the input is faster. For GRIST, all the configurations of SA Fuzzer can trigger 79 bugs. For {\it No H}, {\it $3\times3$} and {\it MNIST}, we reported time of 0.793, 0.721 and 0.767 seconds respectively (please see our replication package for details). 

We further evaluated the impact of SA classifier architecture and training data size on the performance of the \tool. When the soft assertions are trained on 40K data points using \texttt{Random Forest} classifier, our SA Fuzzer detected all bugs with an average time of 1.9182 seconds. Replacing the random forest with a deep neural network (DNN) increased the detection time to 4.1412 seconds, making it 2.16 times slower without improving bug detection. Regarding training data size, the 40K model remained the most efficient, achieving an average time of 4.6637 seconds in the 3 × 3 configuration. Reducing the training data to 15K and 5K led to slower detection times of 5.7662 and 8.4913 seconds, respectively. The 5K model also missed one bug. %highlighting the importance of sufficient training data. These results demonstrate that while a deep learning-based classifier incurs higher computational cost without added benefits, maintaining an adequate training set is crucial for accurate and efficient bug detection.

\section{Threats to Validity and Discussions}~\label{threat}
%We inThe primary potential internal threat to our study's validity lies in the accuracy of our Neuro-Assert Tool%in the correctness of our implementation of Neuro-Assert Tool and oracles generation process. To mitigate this threat, we carefully reviewed the implementation details and utilized the PyCharm IDE's debug mode to verify the correctness of intermediate states throughout the tool's execution. Additionally, we discussed about the criteria for generating oracles. This process was further strengthened through rigorous discussions among the authors and contributors. 
%\wei{done by wei}
When constructing oracles, we confirmed some of the unstable functions with the PyTorch developers. We performed code review and debugging to make sure our implementations of the oracles, the soft assertion generators and the fuzzer were correct. We used 94 ML applications consisting of the programs from GRIST as well as the real-world code from GitHub. We followed a clearly defined procedure (Section~\ref{benchmark}) to select programs to construct our benchmark. We trained soft assertions for 25 unstable functions covering a variety of categories. We used the 5 SOTA baselines for comparison. We ran the fuzzers 5 times to get average results. We also manually inspected the code to confirm the triggered bugs, one of which has been confirmed by the GitHub developers.

To explore the possibility of using traditional specification inference tools to infer safe input conditions for unstable functions, we tried the open-source tool Daikon~\cite{Daikon}.  The first challenge is that Daikon is not compatible with Python-based ML frameworks like TensorFlow and PyTorch. Thus we ported an unstable function \texttt{exp()} to a C++ implementation and ran unit tests to get the traces for Daikon. Daikon failed to report any invariants related to matrix elements; instead, it provides irrelevant properties such as \texttt{return>x} and many other trivial properties.  In this case, the safe input condition is that every element in the input matrix should be less than 88.72 (see our oracle documentation in the replication package). We successfully generated a soft assertion for this function and detected the instability caused by this unstable function in ML applications.

% All the triggered bugs can be confirmed by our oracles.
%One significant external threat is our evaluation subject, GRIST benchmark~\cite{yan2021exposing}. Although GRIST~\cite{yan2021exposing} comprise a total of 79 real-world  programs, the largest collection to our knowledge, inconsistencies across machine learning (ML) library versions limited our analysis to only 19 programs, alongside an additional 11 programs collected from GitHub. Our tool is  developed with the latest version of TensorFlow, which introduces version compatibility issues with GRIST.
% This version compatibility means, as of now, certain programs initially evaluated by GRIST cannot be directly tested with our tool without updates or modifications to align with the current TensorFlow version.

%Moreover, to mitigate bias in the empirical study, we compared the performance of our tool against three Python fuzzers and the SOTA GRIST~\cite{yan2021exposing} for 10 iteration. Although these 30 programs may not fully capture the diversity of real-world application, we plan to evaluate neural-assert on more ML/DL programs in future to alleviate this issue.
\section{Related Work}

\noindent{\it Numerical Bugs}: GRIST~\cite{yan2021exposing} and our approach both aim to detect numerical bugs in ML codes. GRIST is also used automatic differentiation to determine input adjustments. Our novelty is that we used automatically generated soft assertions to guide the input search, while GRIST used manually-encoded safety/error conditions. As a result, we supported much more unstable functions where we cannot manually write safety/error checks. Also, we can find numerical bugs using 6 types of oracles, which allows us to find more types of numerical bugs, while GRIST used NaN/INF to detect failures.
DEBAR~\cite{zhang2020detecting} used abstract interpretation to statically detect numerical bugs. RANUM~\cite{li2023reliability} combines GRIST and DEBAR and fixed the numerical bugs by clipping the values based the range reported by DEBAR.  There are also empirical studies on numerical bugs~\cite{wang2022empirical,di2017comprehensive, vanover2020discovering,jia2020empirical,tambon2024silent,majdinasab2023empirical}. For example, DeepStability~\cite{kloberdanz2022deepstability} studied 
numerical bugs in deep learning libraries. It built a database of numerical bugs, from which, we collected 56 unstable functions. We added more unstable functions and also designed 6 types of oracles and implemented it for 25 unstable functions. %Wang et al.~\cite{wang2022empirical} studied 400 numerical bugs from the PyTorch and TensorFlow libraries and categorizes the bugs into nine types.

\noindent{\it Fuzzing for ML code and models:} Fuzzing has been applied to ML applications~\cite{yan2021exposing}, ML libraries~\cite{tan2019cradle,yang2023fuzzing}, ML models~\cite{Pei2017DeepXplore,odena2019tensorfuzz,You2023DRFuzz,guo2018dlfuzz}, and deep learning compilers~\cite{liu2023nnsmith}. $\Delta$Fuzz~\cite{yang2023fuzzing} applied different execution scenarios as test
oracles to test first- and high-order gradients  targeting the Automatic Differentiation component in DL libraries. CRADLE~\cite{tan2019cradle} used differential testing and found the equivalent functions in the libraries as oracles for fuzzing. DRFuzz~\cite{You2023DRFuzz} found regression faults in ML models by generating inputs that maximize output discrepancies. DeepXplore~\cite{Pei2017DeepXplore} aimed neuron activation coverage to fuzz neural network models. DLFuzz~\cite{guo2018dlfuzz} used mutation and differential testing to test DNN models aiming to achieve neuron coverage. NNSmith~\cite{liu2023nnsmith} used differential testing and gradient based search to find model inputs to test DNN compilers. 
\section{Conclusions}
%As ML code becomes more important and prevalent, there is an urgent need of an automatic fuzzing tool that can quickly and effectively find numerical bugs in the code. 
%\wei{done by wei}
In this paper, we proposed a new concept {\it soft assertion} and developed {\it Soft Assertion Fuzzer} to find numerical bugs in ML applications.  Our work is motivated by the fact that it is very hard, even for ML experts, to write {\tt assert} or symbolic condition checks for numerical stability of ML code, due to the high dimensional input types such as tensors and complex math functions. The soft assertion is an automatically learned ML model that encodes the safety/error conditions for an unstable function.  To generate soft assertions, we first built a database of unstable functions and constructed their testing oracles. We  developed a novel approach of generating the training dataset by running the unit tests of unstable functions. During fuzzing, soft assertions provide signals to guide fuzzers to perform effective and targeted mutations. We performed a comprehensive evaluation on 94 ML applications and compared with 5 SOTA baselines. Our results show that soft assertions are very effective to guide fuzzing. We detected 92 numerical bugs, 13 of them are unknown bugs from the important real-world GitHub code, significantly outperformed the baselines. While all the baselines focus on reporting NaN/INF failures, we found numerical bugs that produced  incorrect output. In the future, we will continue improving soft assertion accuracies and explore more applications of soft assertions.

\section{Data Availability}
Our replication package is located at {~\url{https://figshare.com/s/6528d21ccd28bea94c32}}.

% \section*{Acknowledgments}
% This research was supported in part by XXX grants XXX.
% We would like to thank Ashwin Kallol Joshy for his helpful guidance and support in understanding the project during its early stages. We also thank the anonymous reviewers for their valuable and insightful comments, which greatly improved the quality of this work. The views expressed in this paper are solely those of the authors and reflect the constructive feedback provided by the reviewers.

% We thank the anonymous reviewers for their valuable feedback.
% This research is partially supported by the U.S. National Science Foundation (NSF) under Award $\#2313054$

\begin{acks}
This research is partially supported by the U.S. National Science Foundation (NSF) under Award $\#2313054$. We thank the anonymous reviewers for their valuable and insightful comments, which greatly improved our work. We also thank Ashwin Kallol Joshy for his helpful guidance and support in understanding the project during its early stage.
\end{acks}

%In this paper, we only explored one application of neural assertion. As the widely applicable assertions and invariants,

% explore more applications, e.g., neural assertion aided debugging. We will design more rich labels for neural assertion based on applications

%More types of numerical bugs such as math overflows

%more architecture for an integrated modelpicture tables

\bibliographystyle{ACM-Reference-Format}
\bibliography{main}

\appendix

\end{document}